\documentclass[journal,a4paper]{IEEEtran}
\usepackage{cite}
\ifCLASSINFOpdf
   \usepackage[pdftex]{graphicx}
   \DeclareGraphicsExtensions{.pdf,.jpeg,.png}
\else
   \usepackage[dvips]{graphicx}
   \DeclareGraphicsExtensions{.eps}
\fi
\usepackage[cmex10]{amsmath}
\usepackage{array}
\usepackage{fixltx2e}
\usepackage{dblfloatfix}
\newcommand{\Sec}[1]{Section~\ref{sec:#1}}
\newcommand{\Fig}[1]{Figure~\ref{fig:#1}}

\newcommand{\Eq}[1]{Eq.\ (\ref{eq:#1})}
\newcommand{\Table}[1]{Table~(\ref{tab:#1})}

\begin{document}
\title{The SMARDDA Approach to Ray-Tracing and Particle Tracking}

\author{Wayne~Arter,~\IEEEmembership{Member,~IEEE,}
        Elizabeth~Surrey,~\IEEEmembership{}
        and~Damian.B.~King,~\IEEEmembership{}
\thanks{W.~Arter, E.~Surrey and D.B.~King are employed by the United
Kingdom Atomic Energy Authority, address: CCFE,
Culham Science Centre, Abingdon, Oxon. UK OX14 3DB
e-mail: (see http://www.ccfe.ac.uk).}
\thanks{Manuscript received XXXX.}%
\thanks{Based on a paper presented orally at ICNSP'09~\cite{Wa09f}.}%
}

\markboth{IEEE Transactions on Plasma Science, submitted}%
{Arter \MakeLowercase{\textit{et al.}}: SMARDDA }

\maketitle

\begin{abstract}
There is an increasing need to model plasma interaction with
complex engineered surfaces, notably to verify that power deposition
rates are acceptable. The SMARDDA algorithm has been developed
to meet this requirement, with particular reference to the
neutral beam ducts that feed into the vacuum vessels of tokamaks.
Application to limiters and divertors is made in a
companion paper.
The algorithm is described in detail, highlighting key novel
features, and illustrative duct calculations presented.
\end{abstract}


\IEEEpeerreviewmaketitle

\section{Introduction}\label{sec:intro}
\IEEEPARstart{T}{he} SMARDDA software suite was developed initially in 2007/08
to model the interaction of beams of neutral particles with the low temperature
gas found in neutral beam ducts. The broader context of this work is the 
need to heat using beams of neutral atoms,
high temperature plasmas in magnetic confinement devices.
The prototypical device of this kind is the tokamak,
where the plasma is produced and confined by a multi-Tesla magnetic field in a toroidal
vacuum vessel. The ultimate aim of such
experiments, as exemplified by the multi-billion dollar ITER tokamak currently undergoing construction
in Southern France, is the development of a low-Carbon, electricity power source from
controlled nuclear fusion. Significant amounts of neutral beam power, in the
range of tens of MegaWatts, are presently required in most scenarios for ITER
operation at energy breakeven, and are likely to feature in any tokamak reactor
design.
\subsection{Duct problem}\label{sec:duct}
\begin{figure}[!t]
\centering
\includegraphics[width=2.5in]{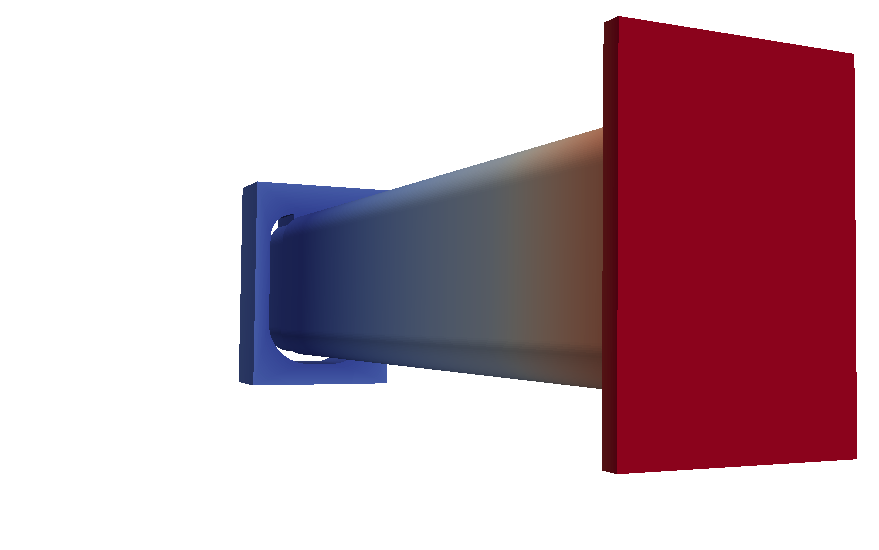}\\
\includegraphics[width=2.5in]{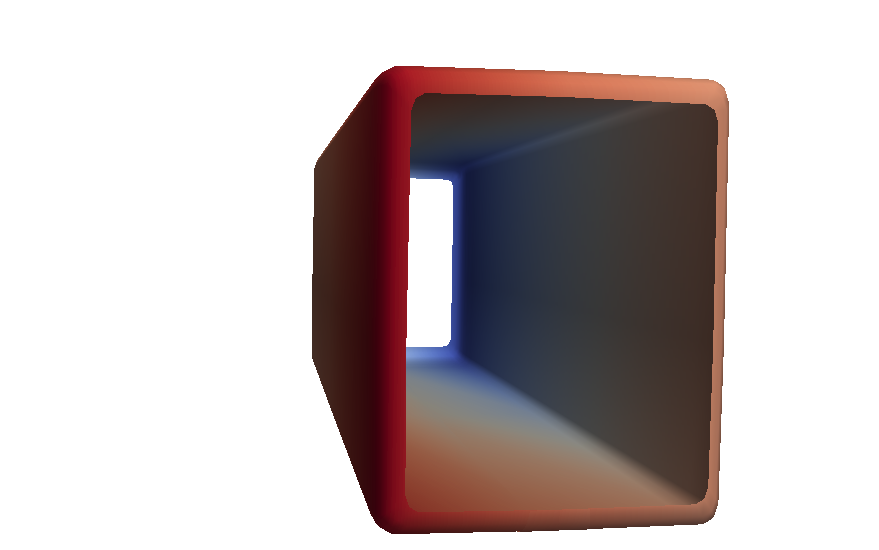}
\caption{Views of ducts. The top shows a duct geometry extracted from
the ITER geometry data, with the end blanked off. The duct
is approximately $5.5$\,m long and $1$\,m wide. The bottom
shows a purely indicative design, indicating the entry and exit.}
\label{fig:ductviews}
\end{figure}

The beams of neutral atoms are produced by a complicated process which
involves first accelerating a beam of ions and then neutralising the
beam particles. Evidently, the acceleration stage must take place at
a distance from the strong confinement fields in the main toroidal
vessel. In a reactor design, there are also good radiological reasons
for wanting complex engineered structures to be remote from the hot
tokamak plasma. Hence the need for neutral beam ducts, see \Fig{ductviews}.
The duct walls will typically be defined using a CAD (Computer Aided Design) package,
and may have a complex design to satisfy a range of different engineering constraints
including the need for surface cooling. This complex geometry is ultimately
represented in CAD geometry database using a special kind of spline
function, known as a NURBS, for non-uniform rational B-spline~\cite{farin}.

Ideally the ducts are narrow enough to minimise the evacuated volume,
yet allow passage of the neutral beam from
their point of production into the central torus without hindrance.
There are typically two sources of gas within the duct, ions
escaping from the tokamak plasma and gas collected in the duct walls that
is released by the beam's impinging upon it.
If the former source dominates, 
gas density will decrease away from the duct's joint with the torus.
Interaction between the beam and the background gas will ionise a fraction of
the neutral beam particles and, under the influence of the tokamak magnetic
field, these particles are likely to impact the duct wall and deposit a
significant amount of heat.
The initial aim of the SMARDDA development is to quantify power deposition
on the wall, both from reionised particles, and from direct impact by
the fast neutrals on the outskirts of the beam.
A companion paper~\cite{Wa14a} describes application of SMARDDA-based
software to the calculation of power deposited by a tokamak plasma
on limiters and divertors.

\subsection{Related problems}\label{sec:related}
There are two main problems to be treated by SMARDDA. The first is the tracking
of charged particles in magnetic fields, corresponding to the motion
of the reionised particles in the stray field in the duct. The second
is to test particle paths for intersection with geometry, corresponding
to the interaction of either the charged or neutral particles with the duct walls.
To get good particle statistics in order to calculate power deposition
accurately, it is desirable to solve these problems in as efficient a
way as possible consistent with the likely accuracy of both the computed field
and the designed surfaces.

Mathematically, the first problem is about the solution of ordinary
differential equations (governing single particle motion) defined
by discrete field samples (meaning the magnetic field is the result of
a separate numerical calculation returning values on a separate grid).
The second problem reduces to the ray-tracing problem, the efficient
calculation of the intersection of a straight line segment with complex geometry.
This is potentially the biggest cause of inefficiency, since the direct approach
of comparing the particle trajectory with every part of the geometry
gets very costly when for example, the geometry is defined
by $100\,000$ small elements and $100\,000$ tracks have to be tested
for intersection.

The above mathematical problems arise in a number of previously
studied situations from which
potential insights into the choice of best numerical algorithm 
for a new software development may be gathered.
The two problems arise in combination for the computation of the efficiency of
electron guns~\Sec{egun} and of the motion of tracer particles
in a fluid flow, although there seems to be very little published
on the latter problem. The second, ray-tracing problem, 
arises by itself both in neutronics, in the calculation of neutron
trajectories through reactor shielding, and most obviously
in computer graphics, in the visualisation of complex objects using 
computer display equipment. Moreover, the computer graphics literature
is generous in publishing details of algorithms, a
distinct advantage for taking ray-tracing techniques
from this field and applying them to others.
\subsubsection{Electron guns}\label{sec:egun}
In gun codes~\cite{He88EGUN}, a charge distribution is
calculated by tracking electrons from emitter to dump, then recalculating the
electron trajectories in the resulting electric field to give a new charge
distribution and so on until a consistent solution is achieved.
In PIC or `particle-in-cell' simulation~\cite{hockneyeastwood,birdsalllangdon}
of electrostatic devices such as electron guns, 
the charged particle paths are normally computed using the ``Boris" scheme~\cite{hockneyeastwood},
a leapfrog scheme which is second-order accurate in both space and time
as well as being symplectic, ie.\ having excellent conservation properties.
Unfortunately, Auerbach and Friedman~\cite{Fr91Nume,Au91Long} have discovered curious
resonance type effects using a fixed timestep. Use of an adaptive scheme,
whereby timestep size is changed to meet accuracy requirements, is however
likely to mitigate the effect.

Electromagnetic field values are calculated by direct product spline interpolation
between grid values of the fields, often a linear interpolation formula is used.
The resulting curved trajectories are typically modelled as sequences of short straight
lines or tracks, each corresponding to one timestep of the Boris particle advance
algorithm.

Directly adopting this approach has several difficulties. The particles
are tracked on the same computational grid which is used for calculating the
fields. To ensure charge conservation, this grid is often a cuboidal lattice,
and material boundaries are forced to coincide with the faces of the
the grid cells or ``voxels" (short for volume pixels). This makes for a
ragged brick-like approximation to the geometry with consequently a
highly inaccurate representation of the surface normals.

More recent PIC work~\cite{As92Part,Ha07Effi} has used unstructured grids, so that vacuum regions
are filled irregularly with tetrahedral elements or ``tets" rather than with uniform, hexahedral, voxels.
The key idea~\cite{As92Part,Se98Adva} is to use local coordinates within each ``tet",
so that particles may be tracked efficiently and accurately across the grid.
This concept was later employed in the CTLASS/MICHELLE software~\cite{Pe05Rece}.
Unfortunately there is still a drawback for present purposes, in that the optimal grid
spacing for the electromagnetic field may be very different from the optimal
size of particle step, for example in the case of neutral particles.

\subsubsection{Neutronics}\label{sec:neuts}
For deep shielding problems
in neutronics, it is critical not to lose any particles due to small, round-off level,
mistakes in the meshing, because the entire radiation flux
may result from as few as three critical particles. Treatment of this
problem, as exemplified by the Monte Carlo N-Particle (MCNP) software~\cite{X503MCNP1},
requires representing the geometry by using the CSG (Constructive Solid Geometry)
representation, ie.\ by intersecting a set of primitive quadric solids such as ellipsoids
and tori. CSG has the advantage that particles may be
located in cells defined by typically a small number of intersecting solids.
This representation enables a `belt and braces'
approach to the movement of particles, because a surface intersected by a track
must form part of the definition of the two cells in which the track starts and ends.
Provided there are no mistakes in the geometry definition, it is plausible
that less than one track in~$10^{12}$ is lost due to round-off errors
assuming a double precision numerical representation accurate to 
$13$ or so decimal digits.
It is unclear that this extreme ability is strictly necessary in duct problems or
indeed many others where the motions of many particles are likely to share similar
properties.

Of relevance to the present study is the ability of MCNP and similar
software to treat particle interactions such as absorption with a background medium.  
This leads to the reionisation algorithm described in \Sec{reion} below.
\subsubsection{Plasma neutral transport}\label{sec:degas}
The DEGAS software~\cite{St94Neut} tracks a gas usually consisting
of thermal neutrals through a volume meshed with tetrahedra.
Unlike the PIC work described above, local coordinates are not used,
but a `belt and braces' like in the neutronics work is employed.
\subsubsection{Computer graphics}\label{sec:graphics}
Computer graphics software must for most purposes operate in real-time,
a typical example's being the ability to spin an illuminated complex
geometrical shape for the user to visualise.
Most algorithms which compute detailed views on computer display equipment operate
by following rays from the user's (virtual) screen to the geometry
and ultimately to a light source~\cite{wattsq}. 
(This technique corresponds to the adjoint approach in neutronics.)
Of the three different problems described in this section, ray-tracing
is easily the most tolerant of error,
since the human eye can easily compensate for an error rate of $1$ in $10\,000$ pixels.

For computer graphics, the importance of speed 
results in a discretisation
of the geometry which is as simple as possible, namely as a set of
triangles to represent all the surfaces in the scene. Ray-tracing
can then be reduced to a vast number of repeats of a very simple problem, namely to
intersect a straight-line segment with a triangle, so that for example
the repetitive parts of the algorithm can be implemented in hardware.

\begin{figure}[!t]
\centering
\includegraphics[width=3.5in]{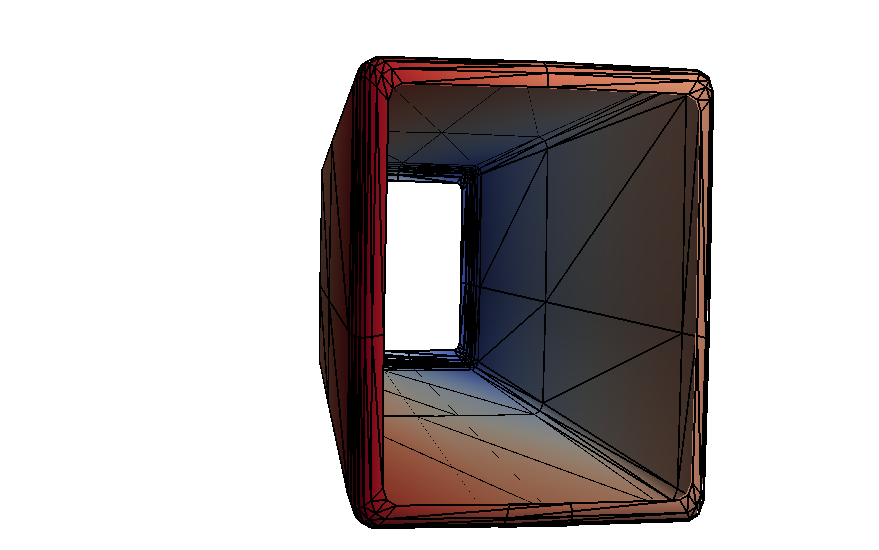}\\
\includegraphics[width=3.5in]{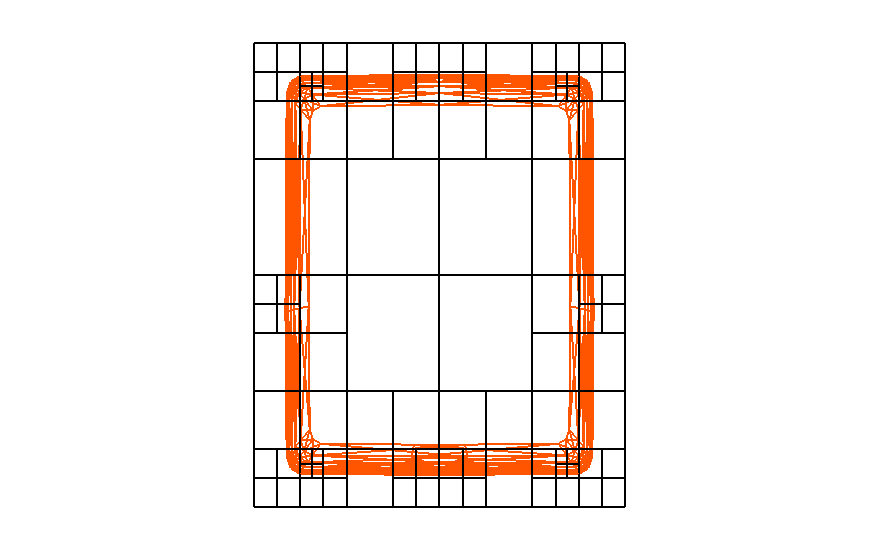}\\
\includegraphics[width=3.75in]{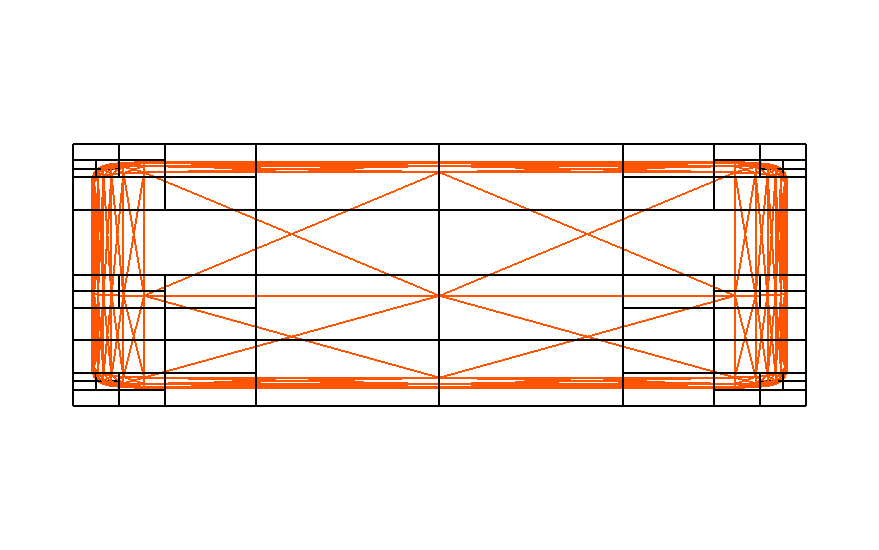}\\
\caption{Mesh of indicative duct design and resulting octree.
The top left shows the mesh projected on the geometry
in perspective, the top right a parallel projection of the mesh,
and octree onto a plane normal to the duct axis.
The bottom is a parallel projection of the mesh and octree onto
a plane at right angles to the first.}
\label{fig:xcelloctmesh}
\end{figure}

Nonetheless it is helpful to reduce the number of triangles~$N_{\Delta}$ which
need to be tested against a particular track. This is achieved by use of
auxiliary data structures, of which there are three main types. The first,
called SEADS (Spatially Enumerated Auxiliary Data Structure) uses the voxel
concept introduced in \Sec{egun}. To each voxel is associated a list of the
surface triangles which intersect it. The idea is that the voxels which
a track crosses may be cheaply found and the track tested only against
those triangles corresponding to the relevant part of the SEADS.
The other two auxiliary data structures by which the triangles are
indexed are hierarchical data structures (HDS),
eg.\ the octree divides the computational domain first into eight equal cuboids,
but then only selectively subdivides these first cuboids into eight, and so on
recursively, depending on details of the scene,
see the projections in \Fig{xcelloctmesh}. Parts of the octree which the
track intersects are identified by a recursive algorithm and then the
corresponding triangles are tested for intersection.
For the HDS in particular, it might be expected that only of
order~$\log_2{(N_{\Delta})}$ objects
need be tested for intersection with a given track.
However, Chang~\cite{chang,chang04} describes how, for specific, selected
distributions of objects in a scene, both SEADS and HDS may still require
$\mathcal{O}(N_{\Delta})$ intersection tests.

\section{SMARDDA Ray-Tracing}\label{sec:RT}
Whilst the mathematical problem of calculating the charged particle motion
can be regarded as solved by use of the Boris algorithm, its requirement
to use a regular mesh is challenging for efficient solution of the
mathematical ray-tracing problem to compute particle collision with the wall.
A further constraint is provided by the need to discretise the NURBS-based
geometry
produced by CAD packages. To minimise effort, this is best achieved by
meshing with pre-existing software, but meshing packages have their own
limitations. There seems to be little software available
to mesh NURBS consistently with a uniform cuboidal lattice, and 
the CSG representation that uses solids is fundamentally unable to treat the
surface representation (``B-rep") normally implied by use of NURBS.
Thus an approach that involves the meshing of B-rep NURBS is
indicated. Demand from the finite element community means that there
a range of software to produce good tetrahedral meshes from NURBS. Since
tet meshers normally commence by triangulating the surfaces, which is
a technically easier problem, this implies a widespread capability to
triangulate NURBS.

The literature reviewed above contains no algorithm which is able
to treat efficiently straight particle trajectories which
are both very long, in the case of the fast neutrals, and relatively
short, in the case of the reionised particle motions over one timestep. 
The need to treat long trajectories
argues against use of tet meshes, since many tets will have to be crossed,
and in favour of working with the surface triangulations.
This motivates the development of a new ray-tracing algorithm which incorporates
many of the ideas from the computer graphics literature.
\subsection{SMARDDA Algorithm}\label{sec:SMARDDA}
The SMARDDA algorithm (pronounced ``smarter") is so named because it represents
a hybrid of two distinct
ray-tracing algorithms which respectively use the octree HDS and SEADS.
The SMART algorithm for ray-tracing using an octree
is described in a paper~\cite{Sp91SMAR}, generally
regarded as difficult to understand, and the DDA
(Digital Differential Analyser) algorithm uses SEADS~\cite{Fu86Arts, Hs92Acce}.
The DDA represents an efficient algorithm to advance a track
long compared to cell size through a SEADS cell-by-cell, 
in much the same way as a charge-conserving PIC algorithm is required to do.
The present section outlines the combined algorithm, mathematical details of key parts of 
SMARDDA are presented in \Sec{samec} and the Appendix. A hybrid of SEADS
and with a different, binary-spaced partition HDS has
been examined by others~\cite{Pr91Adap}.

The dimensions of the smallest cuboid in the octree can be used to ``quantise"
the position of a particle, ie.\ to assign to it coordinates that each are
a multiple of the corresponding cuboid side.
By translating the physical
coordinates if necessary, it may be arranged that the origin of geometry is
close to the zero of the quantised coordinates, so that
all particles within the geometry have positive coordinates, each coordinate
filling a large interval such as~$(0,2047)$.
A key part of the SMART algorithm is the realisation that the binary
arithmetic operation of exclusive-or can then be used to locate particle positions
relative to the octree. In particular, this leads to a simple test as to whether
the positions at the start and end of a track are in the same octree cuboid.

The same-cell test is best illustrated by example, but see
also \Sec{multi}. Suppose say that two of the
particle coordinates have identical integer part, so as to reduce the comparison
to a 1-D problem, and further suppose that the octree cell is 
of size~$2^4=16$. The test is to shift out the 
trailing~$4$ binary digits of each integer-truncated particle position, then apply
the bit exclusive OR function of FORTRAN called {\tt IEOR} on the remaining
bits.
For, consider the binary integer representations of three points
\begin{eqnarray*}
\;1 & = 0...00001 \;\;\mbox{shift}\;\; \rightarrow 0...0 \\
\;7 & = 0...00111 \;\;\mbox{shift}\;\; \rightarrow 0...0 \\
17 & = 0...10001 \;\;\mbox{shift}\;\; \rightarrow 0...1
\end{eqnarray*}
Supposing that the positions of the particle at the track ends are $1$ and~$7$,
shifting the bits and applying {\tt IEOR} as indicated above
gives zero, indicating the
positions occupy the same cell, but not so for the positions~$7$ and~$17$.

The key idea in SMARDDA is that, having constructed an octree to index
the objects in a scene, rays or tracks can be followed through the smallest
cuboids at the deepest levels of the octree in a sequential manner analogous
to the DDA algorithm, as explained further in the Appendix.
An important refinement of the algorithm
is to use a multi-octree structure to treat elongated physical structures
as explained in the next \Sec{multi}.

\subsection{Multi-octree}\label{sec:multi}
\subsubsection{Description}\label{sec:multidesc}
\begin{figure}[!t]
\centering
\includegraphics[width=3.5in]{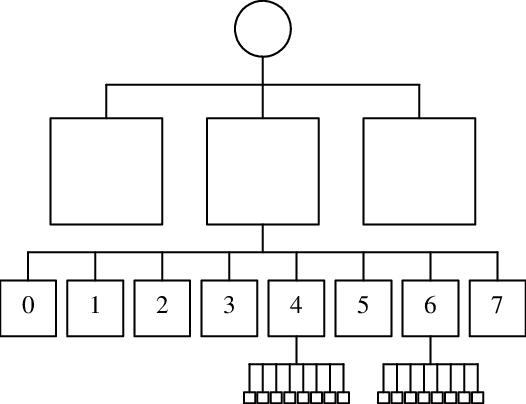}\\
\caption{Schematic of the logical structure of a multi-octree.
Each of the largest squares represents an octree that describes
a congruent cuboid volume, and the further subdivision of one such ``parent"
into eight ``children" and so on is also indicated.
Each successively smaller size of cuboid corresponds to an increasing ``depth" in the tree.
By convention and somewhat confusingly, the ``root", drawn as a circle, 
of the ``tree" is at the top.}
\label{fig:mloctree}
\end{figure}
The octree is a widely used data structure, but it should be evident
that it may become suboptimal for representing duct geometry which
is almost by definition, elongated in one coordinate direction.
This anisotropy has motivated the definition of the multi-octree,
essentially the introduction of an additional hierarchical level
to contain octrees, see \Fig{mloctree}.
Essentially the geometry can be thought of as encompassed by a brick
built out of a set of smaller, identical octree bricks
arranged in a rectangular~$n_x\times n_y\times n_z$ array.
In the case of a duct aligned along
the $y$-axis it is probable that $n_x=n_z=1$ and $n_y$ will approximate
the duct aspect ratio. The size~$L_x\times L_y\times L_z$
of each brick is chosen such that the volume~$n_xL_x\times n_yL_y\times n_zL_z$
does indeed encompass all the geometry.

The next step involves producing an octree indexation of the CAD surfaces
or more precisely their triangulations, within each $L_x\times L_y\times L_z$~cuboid.
This is implemented as a two-stage process. The first stage uses the
classic octree recursion and termination criterion~\cite{wattsq}. 
Cuboids containing triangles are identified~\cite{Mo02Fast} and inspected, and
where necessary the cuboids are divided into eight to reduce the
number of triangles they contain. This process continues
recursively until each contains a maximum specified number~$N_g$ of
triangles, typically $N_g=20$ is found to give good results.
There are the important details that 
\begin{enumerate}
\item Before indexing starts, coordinates are quantised using the
vector~${\bf h}=(h_x,h_y,h_z)$, defined by  $h_i=L_i/2^{N_Q},\;\;i=x,y,z$
where normally $N_Q=10$.
\item It is necessary to recognise the case where further
subdivision does not reduce the number of triangles within a cuboid
(typically this occurs when the cuboid is smaller than the triangles).
\end{enumerate}

Suppose that the above algorithm leads to a depth~$N_O\leq N_Q$ of octree, meaning that
the number of levels in the resulting octree, excluding the `root'
(see the lower part of \Fig{mloctree}) is~$N_O$. 
This octree construct is then revised as follows: every cuboid is
examined to see whether it is
empty or not, and if not, subdivided into eight, but no lower levels
than~$N_O$ are allowed. Empty cuboids are indicated by a nodal marker.
whereas each non-empty cuboid has an associated ``linked list" of
the triangles which intersect it.

\subsubsection{Advantages of the multi-octree}\label{sec:advantage}
\begin{figure}[!t]
\centering
\includegraphics[width=3.5in]{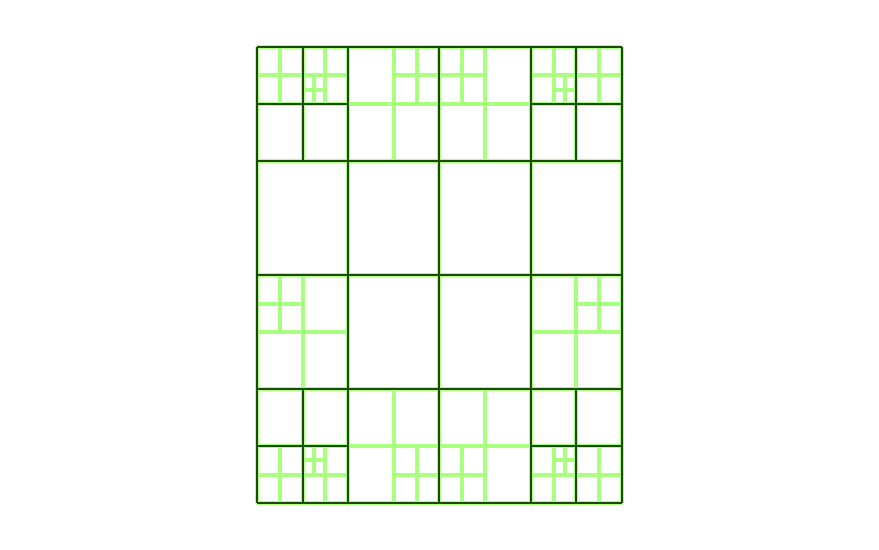}\\
\includegraphics[width=3.5in]{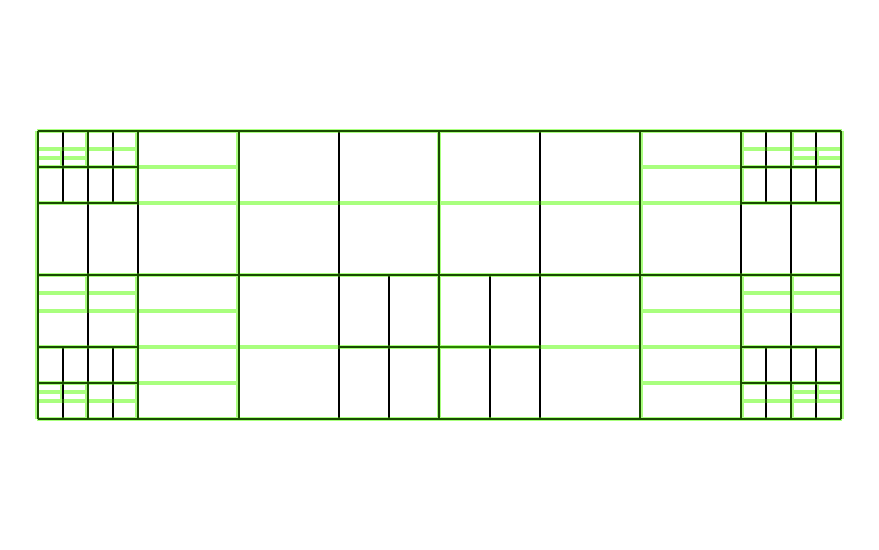}\\
\caption{Comparison of octree and multi-octree structures
for the indicative duct meshing.
The broad pale lines are those of the classical octree,
whereas the narrow heavy lines
indicate a $1\times4\times1$~multi-octree reflecting the proportions
of the duct. The top figure shows
a parallel projection onto a plane normal to the duct axis while
at bottom is a sideways projection
onto a plane at right angles to the first.}
\label{fig:cfoct}
\end{figure}

\Fig{cfoct} shows the octree and multi-octree which result from application
of the above algorithm to a relatively coarse meshing using $448$~triangles
of the indicative duct shown in \Fig{xcelloctmesh}. The standard octree has 
$M_g=456$~cuboidal cells whereas, for
the same parameters, the  $1\times4\times1$~multi-octree has only~$M_g=221$.
For the SMARDDA algorithm generally, this is expected to result in an
approximately  proportionate computational speed-up, but for
application to the duct problem see \Sec{Stats}.

A further advantage is derived from the quantisation of the positions~${\bf x}$.
Suppose the quantised or integer vector corresponding to~${\bf x}$ is
given by ${\bf x}_Q$, so that (possibly following a vector translation)
\begin{equation}
x_Q = x/h_x,\;\;\; y_Q = y/h_y, \;\;\; z_Q = z/h_z
\end{equation}
Suppose further that each of $x_Q, y_Q$ and~$z_Q$ is represented in terms of its bit
representations in a $16$-bit integer word. Assignment
of position in each coordinate to the octree then proceeds as follows.
\begin{enumerate}
\item The first group of bits on the left (most significant bits), 
addresses the octree in which the vector~${\bf x}$ lies, so that
extracting this group from each of $x_Q, y_Q$ and~$z_Q$ gives the
location in~$(n_x,n_y,n_z)$-space.
\item Mask the bit representations to hide this first group, then
form the most significant unmasked bit in each component into a
bit-vector~${\bf x}_q$ that indicates in which
of the eight cuboids at the next level  the vector lies.
\item Descend the tree to the node corresponding to this cuboid, then
use the next most significant unmasked bit in each component to 
descend again and so on until a terminal node or ``leaf"
corresponding to an undivided cuboid is found.
\end{enumerate}
To illustrate point~(2) about the bit-vector, referring to the
node labels $0,1,\ldots,7$ in \Fig{mloctree},  since~$3=011_2$,
node~3  contains the cuboid in position~$(0,1,1)$ relative to the
parent node in $(x,y,z)$~coordinates.

Hence the tree structure may be descended very economically to termination,
by a combination of bit-masking and bit-shifting functions operating
on integers.

\subsubsection{Same-Cell Test}\label{sec:samec}
The movement within an octree described at the end of the preceding
\Sec{advantage} draws heavily from the SMART algorithm~\cite{Sp91SMAR}.
The same ideas may be applied to the problem of determining whether
the two ends of a particle track lay in the same cell. Suppose the 
coordinates of the two
points are truncated to integer values, to give two integer vectors~${\bf Q}^1, {\bf Q}^2$.
Evidently, ${\bf Q}^1, {\bf Q}^2$ and hence the two end-points occupy
the same cuboid of side~$2^{n_{q}}$, if their
bit representations are the same except in the last~$n_{q}$ bits.
Or, equivalently, in more mathematical language
\begin{equation}\label{eq:ieor}
\Sigma_{i=1}^{i=3} ISHFT(IEOR({Q_i}^1,{Q_i}^2),-n_{q}) = 0
\end{equation}
where ISHFT is the bit shift function which is defined so as to move bit
patterns to the right for positive argument, hence the minus sign,
and IEOR is the exclusive-or function introduced earlier.

In many cases, the second end-point will turn out to lie within an adjacent
node, and this case is treated specially for efficiency,
as follows, by computing the vector with components
\begin{equation}\label{eq:ieori}
\sigma_i = ISHFT(IEOR({Q_i}^1,{Q_i}^2),-n_{q})
\end{equation}
(which are anyway needed in the calculation of the sum~\Eq{ieor} above).
If $\sigma_i=1$ for a given~$i$, then the two binary vectors have the same parent
node but they occupy different child nodes, and the index vector~${\bf x}_q$
for~${\bf Q}^2$ differs from that of~${\bf Q}^1$ by the vector with
components $\pm \sigma_i$ where the sign taken is that of~$({Q_i}^2-{Q_i}^1)$.

\section{Duct CAD and Physics}\label{sec:ductphys}
As indicated, SMARDDA expects geometry to be presented as a set of possibly
millions of triangles. To give a well-defined interface the legacy vtk file
format~\cite{vtkusers} is used.
It has the advantage that file contents are easily visualised using the
freely available ParaView software~\cite{paraviewguide}.
As its name implies the legacy vtk file format has
remained unchanged for many years, and is likely to remain current because
a good deal of software, including now SMARDDA, relies upon it.
The question
of CAD to vtk conversion  for the duct problem is now addressed.
\subsection{CAD Conversion}\label{sec:CAD}
The output from a CAD package is seldom suitable for use by physics software.
Invariably, there is too much detail. In the context of duct modelling, only 
surfaces matter and small screws, bolts and other small features
are unimportant provided they
are flush or recessed. However, unintentional small gaps which are insignificant
on an engineering scale may be disastrous in allowing computed rays or
particles to escape the domain. Both these are long-standing issues for 
finite element engineering packages and many options are commercially
available for CAD defeaturing and repair.

The solution adopted relies heavily on the well-known CATIA$^{TM}$ CAD system supplied
by Dassault to perform much of the defeaturing. The CADfix$^{TM}$ package
supplied by ITI TranscenData is largely dedicated to defeaturing and repair,
and can work with CAD from many different vendors, including CATIA files.
Thus it is possible for a user without training in CATIA to make final
adjustments and repairs to CAD produced using CATIA, then use the inbuilt CADfix mesher to produce
a suitable surface mesh of triangles. Locally written software then interrogates
the resulting geometry database and produces the vtk file.
A separate code generates the HDS together with the quantising position
vector transformations.
\subsection{Neutral Launch}\label{sec:launch}
The neutral beam input is modelled as a set of one or more beamlets each
with a Gaussian cross-section. Let $E_b$ denote the energy of particles
in beamlet~$b$ which has a total power of $P_b$.
Supposing without loss of generality
that the beam is directed close to the $z$-direction, then the centre of each beamlet
is specified by co-ordinates~$(x_b,y_b)$ in a plane~$z=z_0$, with its spread given
standard deviations~$\sigma_{xb},\sigma_{yb}$. It follows that particles crossing
$z=z_0$ at position~$(x,y)$ have weights~$w(x,y)$ proportional to
\begin{equation}
P_b \exp{-\left(([{x-x_b}]/\sigma_{xb})^2+
([{y-y_b}]/\sigma_{yb})^2 \right)}
\end{equation}

Neutral particles are launched from $z=z_0$ so as to sample each beamlet.
Quasi-Monte-Carlo sampling is employed,
specifically a 2-D Halton
sequence composed of sequences of vectors with components given  by
Van der Corput sequences of base~$2$ and base~$3$ respectively,
cf.\ the ``quiet start" technique used in PIC codes~\cite{birdsalllangdon}.
Van der Corput sequences contain numbers on the unit interval
generated using the reversed bit patterns of the positive integers~\cite{niederreiter}.
A Halton sequence is completely deterministic from which fact derive
better sampling properties than the more usual Monte-Carlo technique,
yet there is no need to set the length of a Halton
sequence in advance unlike with homogeneous sampling techniques. 
This simplifies the launch of additional particles if additional sampling
is indicated by the results of an initial run.

\subsection{Reionisation and Reionised particles}\label{sec:reion}
The sources of fast ions in the duct are collisions with the background
gas and charge exchange
reactions between the neutral beam and ions in the gas. It will be
assumed for simplicity, and to give a worst case scenario, that
the gas has a uniform density~$n_0$, although the case
of variable density can be treated in a similar way~\cite{Br03Dire}.

To treat reionisation, the preferred approach is to use
particle weighting. The beam particles in any event are weighted by~$w(x,y)$
to reflect the Gaussian distribution of number density in each beamlet,
now the reioinised particles are given a weight~$w_r$ according their
local rate of formation. If the neutral beam particle trajectory
is considered as traversed in $N_{\Delta b}$ small timesteps of duration~$\Delta t_b$, then
\begin{eqnarray} \nonumber
w(x,y,t+\Delta t_b)&=&w(x,y,t)(1 - n_0 \sigma_{ri} v_b \Delta t_b),\\
w_r&=&n_0 \sigma v_b \Delta t_b
\end{eqnarray}
where $\sigma_{ri}$ is the re-ionisation cross-section and
speed~$v_b=\sqrt{2 |q| E_b/m_b}$, with $m_b$~equal to the mass
of beam particles and $q$ is the charge on the electron if $E_b$ is measured
in electron-Volts~(eV).

The above algorithm results in the production
of $N_{\Delta b}$ ionised particles, each assumed to have the velocity
of the impacting neutral. The ions are assumed to travel
to the duct walls without further particle interactions, under the
the Lorentz force law in the static magnetic field ${\bf B}({\bf x})$, viz.
\begin{equation}
\label{e2.1}
m_b \frac{d^2 {\bf x}} {dt^2} = q {\bf v} \times {\bf B},
\end{equation}
where ${\bf x} (t)$ is the particle's position, ${\bf v} = d {\bf x}/dt$
is its  velocity and $t$ denotes time. The numerical details of the
trajectory integration are standard~\cite[\S\,4-7-1]{hockneyeastwood}
given a definition of~${\bf B}$ in terms of samples on a uniform
rectilinear grid. To test for wall collisions, the ion trajectory is assumed
to consist of short straight tracks joining the positions~${\bf x}$
at the start and end of each timestep.


\section{Results}\label{sec:results}
The ITER duct of \Fig{ductviews} is converted as described in \Sec{CAD}
to give a vtk file with $2\,146$~triangles that represents the surfaces.
Since the imported duct geometry is so elongated in the $x$-direction,
this is an interesting configuration for testing the effect of
variations in the $(n_x,n_y,n_z)$-size of multi-octree. Note that
a $(2,2,2)$-multi-octree corresponds to the classical octree.
It was rapidly discovered that it was not possible to produce
octrees of depth ten or less without setting $N_g=100$ or so,
apparently because of the inhomogeneity of the surface meshing.
However, comparatively simple multi-octrees with $200$--$400$
cells were produced with
a maximum of $80$--$100$ triangles per cell, see \Table{Stats} in \Sec{Stats}.

\subsection{Ray Statistics}\label{sec:Stats}
\begin{figure}[!t]
\centering
\includegraphics[width=3.5in]{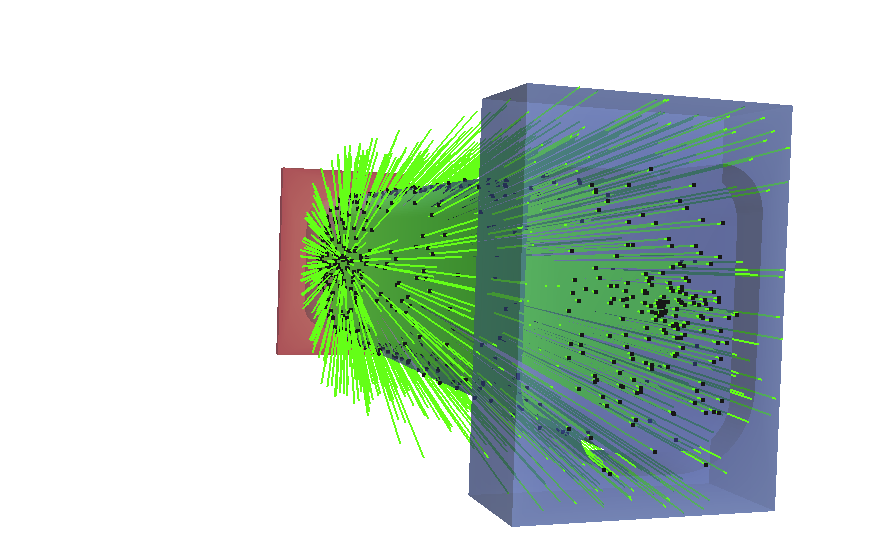}\\
\includegraphics[width=3.5in]{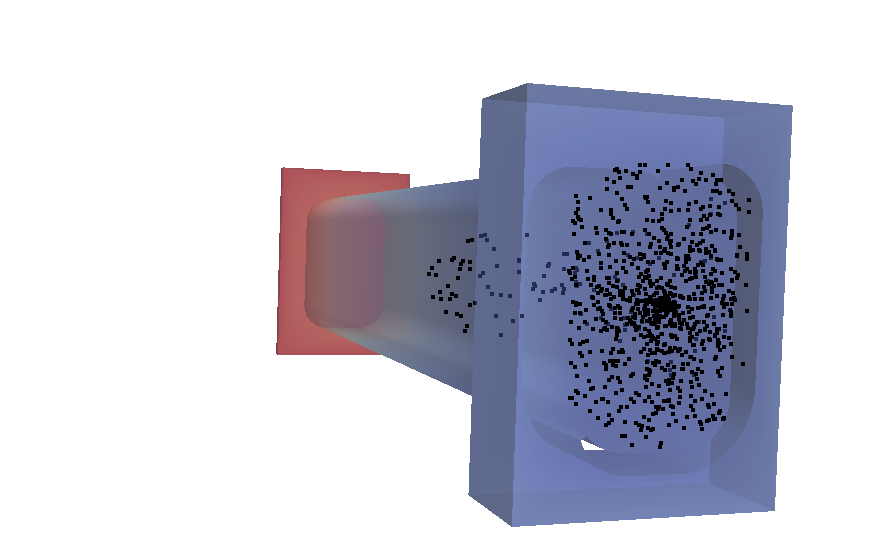}\\
\caption{
Track-geometry intersection tests for
$1000$~fast neutrals in
cones of angle $\alpha=0.5$ (top) and $\alpha=0.1$ (bottom).
The tracks are shown in green, emanating from
the duct centre at rear in upper figure.
The black dots mark the intersections between the tracks and
the duct geometry, or the closure plate at near end.}
\label{fig:Stats}
\end{figure}

As part of the testing process, a ``beam" of neutrals was arranged
to be launched from a single point in the duct centre, to fill
uniformly a cone of angle~$\alpha$ about the long duct axis. Two sizes of~$\alpha$
were employed, $\alpha=0.1$ corresponding to a spread in degrees
of approximately~$5.7$, and $\alpha=0.5$ giving $28^o$ of spread.
The smaller spread is chosen to be indicative for a neutral beam which
fills the duct, the larger spread is designed to generate a significant
number of collisions with with the duct walls as
undergone by the fast ions which strike and heat the wall, see \Fig{Stats}.

Each of the tests recorded in \Table{Stats} involves the launch
of $1\,000\,000$ particles. Hence it will be seen that each
track intersection test takes of order a few micro-seconds.
The variation in cpu time with HDS is significant. It is evident that
the fastest tracking occurs when the number of octree cells
\emph{along} the duct is least. This is most evident when
the tracks are more nearly parallel to the duct for $\alpha=0.1$.
Placing more cells along the duct is evidently beneficial
when tracks are directed more across the duct as in the $\alpha=0.5$ case.

No particle tracking algorithm is perfect, because ultimately 
round-off error causes in the duct problem,
some tracks to miss geometry they should have intersected.
For the cases described in \Table{Stats}, anomalous tracks
were detected by visual inspection
of the plot of end-points each drawn with a large size. When
$\alpha=0.1$, a single anomalous track was detected for the octree HDS,
but all tracks were computed successfully for the other HDS. The
wider beam is more demanding test of the surface intersection algorithm,
but \Table{Stats} shows that the anomalous loss rate is less than
$1$~in~$10^5$ for most cases. This is probably a good rate for a
computer graphics-based algorithm which could easily lose one track in
$10\,000$ without the viewer's noticing. In fact from the distribution
of anomalous points, it is clear that they are associated with octree
faces and much smaller anomalous losses might be easily achievable.
However, the present loss rates are compatible with very accurate
calculations of power deposition in the duct.

\begin{table}[!t]
\caption{This shows the effect of varying the size of~$(n_x,n_y,n_z)$ of
multi-octree on cpu time for the simple neutral beam test. $M_g$ is the number
of octree cells, and the cpu times are for the indicated beam spreads
(in radians).
The last column lists the number of anomalous tracks for $\alpha=0.5$.}
\label{tab:Stats}
\centering
\begin{tabular}{|c|c||c|c|c|}
\hline
$n_x,n_y,n_z$ & $M_g$ & cpu-$0.5$ & cpu-$0.1$ & Lost\\
\hline
$2,2,2$ &  $209$ &  $5.7$ & $4.7$ & $6$  \\
$8,1,2$ &  $366$ &  $23.2$ &  $36.4$ & $17$  \\
$1,8,2$ &  $201$ &  $12.7$ &  $11.1$& $4$  \\
$1,2,2$ &  $245$ &  $5.6$ &  $3.9$& $4$  \\
\hline
\end{tabular}
\end{table}
\subsection{Power deposition}\label{sec:Powerdep}
The power deposited in the ITER duct of \Fig{ductviews} may be calculated
using SMARDDA. Representative parameters are assumed for the neutral
beam, viz. total power of~$20$\,MW and $E_b=1\,000$\,keV. The beam is
assumed to consist of a single Gaussian beamlet centred in the
middle of the duct with $\sigma_x=0.216$\,m,
$\sigma_y=0.338$\,m. 
A high background gas density~$n_0=5 \times 10^{16}$\,m$^{-3}$,
approximating that of the tokamak edge is assumed, and is assumed
to present a cross-section for re-ionisation~$\sigma_{ri}=3.8 \times 10^{-21}$\,m$^2$
to the beam. A representative magnetic field is applied that reaches
approximately~$4$\,T at the torus end of the duct. 

\begin{figure}[!t]
\centering
\includegraphics[width=3.5in]{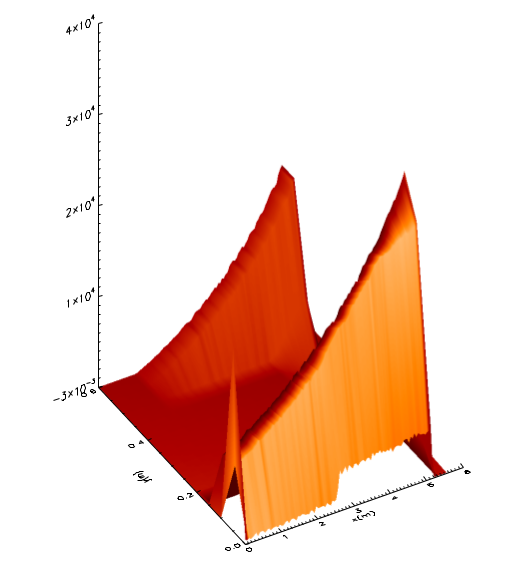}\\
\caption{Power in~$W m^{-2}$ deposited on the top surface of the ITER duct
as calculated using SMARDDA algorithm.}
\label{fig:slide22}
\end{figure}

The beam is sampled using $100$~particles and the trajectory of
each neutral generates~$N_{\Delta b}=300$ ions, so that the total
number of particles followed is~$60\,200$ requiring $12\,040\,000$
tracks to be tested for geometry intersection.
No aberrant particles
are found and the computation completes in under an hour on a laptop,
ie.\ a cpu cost of~$0.25$\,ms/track. The resulting power deposition
profile for the top of the duct is shown in \Fig{slide22}.
The results are easier to interpret than those from older software
which takes a week of cpu~\cite{Wa09f}.
\section*{Acknowledgment}
This work was funded by the RCUK Energy Programme grant number EP/I501045
and the European Communities under the contract of Association between
EURATOM and CCFE.
To obtain further information on the data and models underlying this
paper please contact PublicationsManager@ccfe.ac.uk.
The views and opinions expressed herein do not necessarily reflect
those of the European Commission.

\appendix [Details of the SMARDDA Algorithm]
The SMARDDA algorithm is encapsulated as a function which takes
a straight particle track,
specified by its start and end points, and determine whether
it intersects any part of the geometry. If so, it returns
the first point of intersection.
The track is defined by two arguments of type {\tt posnode},
a type which consists of  a position vector and the identifier of the 
deepest corresponding node in the octree. Frequently, the node
of the start position is known as the result of a previous
calculation, if not, as occurs typically on the first step,
it is calculated by binary look-up in the octree. The second argument
will acquire a nodal value only upon exit, corresponding to
the end of the track. Its position vector will be changed from
the input value only if the track intersects the geometry as
signalled by the return of another, the last, function argument.

\begin{figure}[!t]
\centering
\includegraphics[width=3.5in]{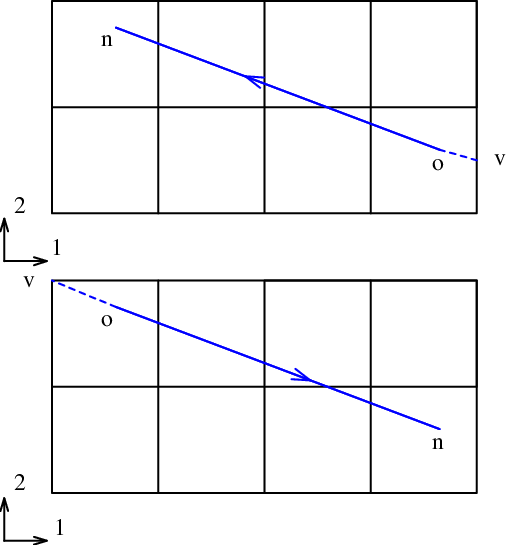}\\
\caption{The track projected on the $x_1 x_2$-plane, with integer
coordinate values shown as a grid.
The track is defined by the start
point~$o$ and the end point~$n$, joined by an arrow
to indicate direction of travel. The virtual origin~$v$
is constructed as indicated to have an integer~$x_1$ coordinate.}
\label{fig:vpos}
\end{figure}

The function uses the same-cell test of \Sec{samec}
as a preliminary check. This is efficient when
particle trajectories are short.
For longer tracks, the
particle advance is controlled by a loop in the first coordinate~$x_1$.
For reasons which will become apparent,
particle track is started from a virtual origin,
viz. an integer (quantised) position in the first coordinate, either the largest
integer less than the start position if the motion is forwards,
or the smallest integer greater than the start position if motion
is backwards, see the illustration in \Fig{vpos}.

In the DDA applied conventionally to a SEADS, each
subsequent increment of the integer loop counter gives
a position in a different cell along the track.
Supposing that lateral motion of the particle relative
to the first coordinate direction is negligible,
the first position coordinate is incremented by one
when motion is forwards, or decremented by unity when motion is backwards. 
First, the track is tested against objects
in the cell indexed by the octree node corresponding to the start position,
and if there is a collision ``in the node", the track
terminates at the intersection point. If there is no collision, there
is a same-cell test of the new position and the end-point of the track.
If this is true, the loop terminates,
otherwise the position is updated and the loop continues in
a new cell with collision tests etc.

The advance along the track is complicated in higher dimensions
because the track may enter new nodes as a result of its motion
in the other coordinates. It is efficient to arrange so that the 
particle moves farthest in the direction of the first coordinate, by
relabelling if necessary. The other two coordinates may be chosen purely
so that the relabelled coordinates~$(x_1,x_2,x_3)$ form a right-handed set. 
In the resulting (quantised) coordinate system, the vector of
advance is $(\alpha_1,\alpha_2,\alpha_3)$ where $\alpha_1=\pm 1$,
$|\alpha_2|<1$ and $|\alpha_3|<1$.
In outline, coordinate~$x_2$ is incremented by~$\alpha_2$ to see
whether this point lies in a new node, then 
coordinate~$x_3$ is incremented by~$\alpha_3$ and this point
tested, and finally
all three coordinates are updated, and the new point tested.
The latter point now on the track, is then updated like the first virtual point,
and so on.

\begin{figure}
\centering
\includegraphics[width=2.75in]{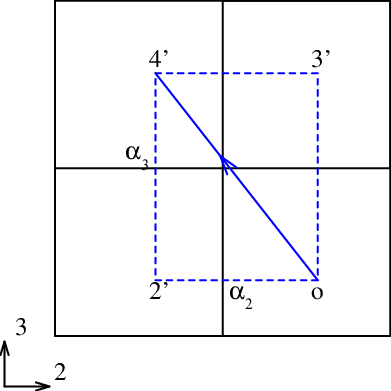}\\
\caption{Segment of track projected on the $x_2 x_3$-plane, with 
boundaries of the smallest local cells shown. As the track is
incremented by a unit quantity in the $x_1$~direction, it moves
from the point~$o$ in the lower right-hand node to the point~$4'$
in the upper-left by the vector displacement~$(\alpha_2,\alpha_3)$. The nodes containing points~$2'$ and~$3'$ are
tested for colliding objects, the former unnecessarily in this instance.}
\label{fig:detupd}
\end{figure}

In detail, consider first the $x_2$~update.
There is a quick test as to 
whether the integer part of $x_2$ has changed,
then if so, a test whether the new~$x_2$ position, shown as
point~$2':(x_1,x_2+\alpha_2,x_3)$ 
in \Fig{detupd}, lies in the current node. If $2'$ is found to 
lie in a new node, the objects in that node are tested for
collision with the track, and if no collision is found, the code proceeds. The
coordinate~$x_3$ is updated to give the
point~$3':(x_1,x_2,x_3+\alpha_3)$, which of course may lie in a new node,
and if so, the node is tested for collisions in the same way. There
is the additional complication that the
point~$4':(x_1,x_2+\alpha_2,x_3+\alpha_3)$ may lie in a node
different to that of either of the points~$2'$ and~$3'$.
If so, a further test for collision is performed in this third node.
However, this is the maximum number of nodes that
need testing between updates of~$x_1$.

Moreover, and distinguishing
the SMARDDA algorithm from the DDA variant above,
the increment in unity of the
position~$x_1$ is replaceable by increments which
could be as large as $2$,~$4$, $8$~or more.
The idea is to jump along the track as far forwards in the current
cell as possible, to an integer point $x_1$, such that the
next integer increment of $x_1$ will produce a point in a new cell,
possibly laterally displaced.

The underlying mathematics are as follows. For DDA it is anyway convenient
to introduce binary markers $d_j$ which are unity for forwards
motion in direction~$j$ and zero for backwards. Supposed the quantised size of
the cell is $2^{\ell_k}$ where $k$~labels the octree level, then
if $o_j$ are the coordinates of the current cell's origin,
the path increment~$\Delta$ can be calculated using
\begin{equation}
\Delta_j=\left(o_j+ ISHFT(d_j, \ell_k) -x_j\right)/\alpha_j,\;\;\;j=1,2,3
\end{equation}
as
\begin{equation}
\Delta= INT\left(\min{\{\Delta_1, \Delta_2, \Delta_3, |x^E_1-o_1|,
ISHFT(1, \ell_k)-1\}}\right)
\end{equation}
(The last term corresponds to a displacement limited by the cell size,
and the second-last
accounts for a track ending at~${\bf x}^E$.)
One step of the DDA algorithm can then be applied as though beginning from
the position
\begin{equation}
{\bf x}'=(o_1,x_2,x_3)+(\alpha_1,\alpha_2,\alpha_3)\Delta
\end{equation}
but note the need for special treatment at the track end-point.
The concept of virtual origin is important here, in that
the track can be tested in sections defined by a sequence of
virtual origins each corresponding to an integer value of~$x_1$.

Last details are that the testing of the track for collisions is
performed using Badouel's algorithm~\cite{Ba90Effi}, modified to take
into account finite precision computer arithmetic as in ref~\cite{Ha07Effi}.
Where the point of intersection of the track with the object face is
needed, eg.\ for diagnostic purposes, it is computed as in the second step of 
Badouel's algorithm~\cite{Ba90Effi,Ha07Effi}.




\end{document}